\documentclass{article}
\usepackage{ijcai26}

\usepackage{times}
\usepackage{soul}
\usepackage{url}
\usepackage[hidelinks]{hyperref}
\usepackage[utf8]{inputenc}
\usepackage[small]{caption}
\usepackage{graphicx}
\usepackage{amsmath}
\usepackage{amsthm}
\usepackage{booktabs}
\usepackage{algorithm}
\usepackage{algorithmic}
\usepackage[switch]{lineno}

\usepackage{subcaption}
\usepackage{xcolor}
\usepackage[acronym]{glossaries}
\makeglossaries
\newacronym{ctde}{CTDE}{Centralized Training, Decentralized Execution}
\newacronym{knn}{kNN}{k-Nearest Neighbors}
\newacronym[plural=MOEAs, longplural=Multi-Objective Evolutionary Algorithms]{moea}{MOEA}{Multi-Objective Evolutionary Algorithm}
\newacronym{moead}{MOEA/D}{Multi-Objective Evolutionary Algorithm Based on Decomposition}
\newacronym{moqd}{MOQD}{Multi-Objective Quality-Diversity}
\newacronym{nsgaii}{NSGA-II}{Non-Dominated Sorting Algorithm II}
\newacronym[plural=POIs, longplural=Points of Interest]{poi}{POI}{Point of Interest}
\newacronym[plural=POIs, longplural=Points of Interest]{pois}{POIs}{Points of Interest}
\newacronym{qd}{QD}{Quality-Diversity}
\newacronym{spea2}{SPEA2}{Strength Pareto Evolutionary Algorithm 2}
\newacronym{modecpomdp}{MODec-POMDP}{Multi-Objective Decentralized Partially Observable Markov Decision Process}
\newacronym[plural=EAs, longplural=Evolutionary Algorithms]{ea}{EA}{Evolutionary Algorithm}
\newacronym{mome}{MOME}{Multi-Objective Map-Elites}
\newacronym{rl}{RL}{Reinforcement Learning}
\usepackage{amssymb}
\usepackage{placeins}

\title{Entropy-Augmented Multi-Objective Policy Optimization in
Multiagent Systems}

\author{
Jamie Santos$^1$
\and
Ayhan Alp Aydeniz$^1$\and
Raghav Thakar$^1$\And
Kagan Tumer$^1$\\
\affiliations
$^1$The Collaborative Robotics and Intelligent Systems (CoRIS) Institute\\
Oregon State University, Corvallis, Oregon, USA\\
\emails
\{santjami, aydeniza, thakarr, kagan.tumer\}@oregonstate.edu}

\begin{document}

\maketitle

\begin{abstract}

Autonomous agent teams deployed in settings such as marine and extraterrestrial outposts must coordinate actions to achieve optimal outcomes across multiple competing objectives. Multi-objective evolutionary algorithms such as NSGA-II optimize for diversity in the objective space, but neglect diversity in the behavior space, possibly leading to premature convergence and a collapse in behaviors that may differentiate policies in different external conditions. To address this, we introduce an entropy-augmented policy evaluation strategy that incorporates an entropy bonus into agent fitness scores, discouraging behavioral homogeneity across the evolving population. By augmenting policy evaluation with a behavior-space diversity signal while preserving the underlying Pareto optimization framework, our method is designed to encourage exploration of behaviorally distinct policies in multiagent domains. We evaluate our approach across rover-domain experiments with qualitatively distinct reward structures and observe hypervolume improvements of up to 48\% relative to the NSGA-II baseline, suggesting that behavioral diversity is a promising and underexplored direction for improving multi-objective multiagent evolutionary optimization.
\end{abstract}

\section{Introduction}
Robot teams have been increasingly deployed in settings such as marine monitoring, space exploration, and disaster response. These teams must often balance competing objectives such as energy usage, task completion, and team safety. Since a single policy rarely optimizes all objectives simultaneously, as these are often in conflict, learning a set of Pareto-optimal team policies is frequently necessary. Generating a robust set of such policies that provide optimal trade-offs across objectives is therefore crucial for supporting a variety of operator preferences and evolving mission conditions.



\glspl{moea} such as \gls{spea2}, \gls{moead}, and \gls{nsgaii} optimize sets of Pareto-optimal control policies by maintaining diversity across candidate solutions in the objective space during population-based searches \cite{zitzler2001spea2,zhang2007moea,deb2002fast}. Diversity is typically enforced through mechanisms such as crowding distance, which preferentially retains solutions that are more spread out in the objective space. \gls{moqd} approaches extend this idea by optimizing a grid of Pareto fronts across predefined behavioral niches, explicitly accounting for behavioral diversity alongside objective performance \cite{pierrot2022multi,janmohamed2025multi}.


However, these approaches do not integrate behavior-based diversity into the search for the global optimal Pareto front. As a result, behaviorally distinct policies that yield similar objective scores early in optimization may be prematurely discarded. This risk is especially acute under sparse rewards, where behavioral differences are not yet reflected in objective values. For example, rovers learning to visit \glspl{poi} may receive identical rewards whether most of the team sits idle near one \gls{poi}, or whether most of the team wanders near other \glspl{poi}, yet the latter strategy may be closer to achieving higher rewards in fewer generations. In \gls{qd} approaches, behavior is more explicitly accounted for by optimizing Pareto fronts across behavioral niches, but localized competition can suppress policies that can otherwise emerge as globally dominant.

In this work, we introduce a behavior-space-aware fitness-shaping strategy that incorporates behavioral diversity into global multi-objective policy optimization for teams of agents, built on the widely adopted NSGA-II algorithm. Our key idea is to augment the rollout rewards for each objective with an entropy bonus, measured as the estimated average \gls{knn} entropy of joint agent positions across a policy's trajectory. Rather than treating entropy as a separate objective, it is added directly to each objective's reward and scaled by a tunable parameter $\beta$, such that higher values of $\beta$ favor policies that promote broader exploration during evolution. This approach preserves the global Pareto structure while encouraging behavioral diversity throughout the evolutionary search.

We evaluate our approach in the multi-rover domain across experiments with qualitatively distinct reward structures: an obstacle navigation task requiring agent coordination to obtain higher-valued rewards, and a temporal task in which agents must trade off between near-term and long-term rewards. In these experiments, entropy augmentation yielded an average of 18\% higher hypervolume in Experiment 1 and 48\% higher hypervolume in Experiment 2 compared to the NSGA-II baseline. These findings provide preliminary evidence that incorporating behavioral diversity into fitness evaluation can improve multi-objective optimization performance in some setting and motivate further investigation of behavior-aware evolutionary search. While the presented results are encouraging, additional evaluation over more trials and a broader range of domains is necessary to determine robustness and generality of the proposed strategy.

\section{Background and Related Work}

\subsection{Multiagent, Multi-Objective Learning}
Many tasks, such as cooperative ground exploration and mapping and multi-robot search and rescue are naturally multi-agent, multi-objective tasks. For example, autonomous extraterrestrial rover teams may need to work together to efficiently cover ground for exploration while minimizing completion time and energy consumption \cite{nayak2025multi}. In this setting, 2+ agents are tasked with cooperating in a common environment to achieve several shared goals. Furthermore, agents are expected to act simultaneously and independently, and may not be privileged with global information \cite{yamauchi1998frontier}. An example of where this may occur is in underwater robotics where the properties of water make communication between robots especially difficult \cite{heidemann2006research}.

Formally, the multi-objective multiagent setting is modeled as a \gls{modecpomdp}, defined by the tuple $(\mathcal{S},\mathcal{A},\mathcal{O}, \vec{r},T)$ \cite{felten2024momaland}. In the \gls{modecpomdp}, $\mathcal{S}$ represents the state space, $\mathcal{A}=a_1 \times a_2\times ...\times a_i$ the joint action space across all $I$ agents, and $T$ the state transition probability function $\mathcal{S} \times\mathcal{A}\times{\mathcal{S}} \rightarrow[0,1]$. The vector $\vec{r}$ is the immediate reward vector of dimension $n$, the number of objectives. Actions are taken according to each agent $i$'s local observation space, $o_i \in O$, and a decentralized policy, $\pi(a_i|o_i)$, in an effort to maximize the total accumulated  reward vector over the episode horizon
\cite{roijers2013survey}.

Because agents cannot access global state information during deployment, a common approach to solving multi-agent MDPs is to use a \gls{ctde} framework. Using \gls{ctde} in team-based scenarios mitigates issues that arise from relying on shared intelligence during deployment, such as when individual robots fail \cite{yamauchi1998frontier}.

The team is also faced with a dilemma; optimizing returns on one objective may come at the cost of diminishing the outcomes of other objectives. Therefore, there may not be a single optimal solution. Instead, the optimal set of policies in which a single policy cannot be considered objectively better than another, also known as the Pareto front, is the final output \cite{zitzler2004tutorial}. In the multiagent setting, a single solution along the Pareto front often correlates to a joint-team control policy, such as a set of weights for agents' neural network controllers \cite{thakar2025multiagent}. Points along the Pareto front in these cases therefore represent tradeoffs between different controllers, where points along different objective axes represent policies that specialize in achieving specific objectives.

\subsection{Evolutionary Approaches to Multi-Objective Learning}
A classic approach to multi-objective optimization problems is to scalarize the reward by taking a weighted sum of the reward vector $\vec{r}$ (or expected value vector $\vec{V}^\pi$ in reinforcement learning approaches): 
\begin{equation}
\tilde{r}_t = \vec{w}\cdot\vec{r}_t = \sum_{i=1}^nw_ir_{t,i}
\label{eq:scalarization_reward}
\end{equation}
\cite{van2014efficient}.
While simply hand-crafting a vector of importance weights for each objective is sometimes an effective strategy, doing so assumes in-depth knowledge of the relative importance of objectives and that a single fixed tradeoff is sufficient  \cite{tan2005multiobjective}. Furthermore, weighted-sum methods fail to capture the full portfolio of possible solutions in non-convex problems \cite{roijers2013survey,pappas2021multiobjective}.

\glspl{ea} evolve a population of solutions to an objective function. By extension, \glspl{moea} are an established method of optimizing and maintaining a set of solutions over a multi-dimensional objective space. They aim to minimize functions of the form 
\[
F(x) = (f_1(x), \ldots, f_m(x))^T \quad \text{s.t.} \quad x \in \Omega
\] in which $\Omega$ represents the decision space, $x$ a decision vector, $m$ the number of objective functions $f_i: \Omega \rightarrow R$, and $R^m$ the objective space \cite{zhou2011multiobjective}. The set of non-dominated solutions is what is known as a Pareto front; a solution $x$ is said to dominate solution $y$ when
\[\forall i \in \{1,\ldots,m\}, \; f_i(x) \ge f_i(y)\]
and
\[\exists j \in \{1,\ldots,m\} \text{ such that } f_j(x) > f_j(y),\]
i.e., $x$ performs at least as well as $y$ across all objectives, and there exists at least one objective in which $x$ outperforms $y$ \cite{hu2023mo,zhou2011multiobjective}.

Strength-based methods such as \gls{spea2} seek to preserve an even distribution of solutions in the objective space by ranking solutions according to how many others they dominate \cite{zitzler2001spea2}. Objective-space diversity is maintained via \gls{knn} distance. In contrast, dominance-based methods such as \gls{nsgaii} seek to preserve an even spread of solutions across the Pareto front \cite{deb2002fast}. \gls{nsgaii} ranks solutions primarily by Pareto dominance, and then by crowding distance, in each evolution generation. 

However, as the dimensionality of the objective space increases, i.e., 4 or more objectives, solutions become sparser and crowding distance becomes a less relevant strategy for ranking solutions \cite{yuan2014improved}. Reference-based methods such as NSGA-III address this issue by selecting non-dominated solutions nearest to supplied reference vectors \cite{deb2013evolutionary}. Yet, this performance comes at the cost of additional computational overhead. Therefore, for this work, NSGA-II was selected as the baseline in order to demonstrate improved Pareto optimization in lower dimensions.

These methods are primarily designed to optimize diversity in the objective space, but neglect to directly consider the behavior space \cite{lehman2011evolving}. However, evolutionary algorithms themselves are agnostic to what exactly they are evolving, and modifications to the fitness function(s) enable behavioral diversity in solutions, as well. Quality-diversity methods such as \gls{mome} aim to address behavioral diversity by optimizing Pareto fronts across a grid of niches, producing an array of behaviorally distinct Pareto fronts \cite{pierrot2022multi}. While this method explicitly accounts for behavioral diversity, competition occurs primarily within behavioral niches rather than across a single global Pareto front. Furthermore, while MAP-Elites has been extended to both multi-objective \cite{pierrot2022multi} and multiagent \cite{ingvarsson2023mix} settings, combining these ideas remains relatively unexplored. This paper introduces a method of adapting \glspl{moea}, namely \gls{nsgaii}, to consider behavioral diversity as well as objective-space diversity when optimizing the global Pareto front in multiagent settings.

\subsection{Entropy-Driven Behavioral Diversity}
Behavioral diversity has been explored in evolutionary and reinforcement learning as a means to encourage exploration to mitigate premature convergence. Lehman et al. demonstrate that the objective function itself may misguide the policy optimization process, and that searching for behavioral novelty instead can outperform objective-based optimization \cite{lehman2011abandoning}. This idea has been extended to multiagent settings; novelty search has demonstrated that rewarding behavioral diversity can encourage exploration and higher team rewards in single-objective cooperative tasks \cite{aydeniz2023novelty}.

In population-based search settings, entropy is a well-established measure of diversity \cite{haarnoja2018soft}. Maximum-entropy \gls{rl} methods incorporate entropy directly into the objective function:
\begin{equation}
    J(\pi) = \sum_{t=0}^{\infty} \gamma^t \mathbb{E}_{(s_t,a_t)\sim\rho^\pi}\left[r(s_t,a_t) + \beta \mathcal{H}(\pi(\cdot|s_t))\right],
\end{equation}
where $s_t$ represents the states at time $t$, $a$ the action, $r$ the reward signal, $H$ the expected entropy of the policy, and $\beta$ the temperature parameter. $H$ is calculated as
\begin{equation}
    H(\pi(\cdot|s_t)) = \mathbb{E}_{a_t\sim\pi(\cdot|s_t)}\left[-\log\pi(a_t|s_t)\right]
\end{equation}
\cite{dong2025maximum}. 

Entropy can also be estimated over agent trajectories to quantify behavioral diversity across a population. A common metric used is \gls{knn} entropy, which approximates the entropy of a continuous distribution using a finite set of samples \cite{kozachenko1987sample}. Analogously, the maximum entropy approach incorporates entropy directly into the objective function; this work proposes introducing entropy directly to the fitness scores of candidate policies in evolutionary learning methods. This work investigates augmenting evolutionary fitness scores using entropy estimates over agent trajectories. By incorporating behavioral diversity directly into the fitness evaluation, the proposed approach seeks to differentiate behaviorally distinct policies without introducing behavioral diversity as an additional optimization objective.

\section{Method}
In this section we discuss the problem formulation in the multiagent, multi-objective setting and the incorporation of the entropy metric into the fitness evaluation for the \gls{nsgaii} algorithm.

\subsection{Environment: Multi-Agent Rover Domain}
The Rover Domain is a robot-based simulation benchmark for evaluating multi-agent policies under a shared global reward \cite{tumer2002learning,aydeniz2023novelty,thakar2025multiagent}. In this environment, teams of agents are tasked with visiting \glspl{poi} and are rewarded according to the task's specific conditions. Objectives can be defined according to \gls{poi} type, or through other metrics such as time or distance covered. Each rover is equipped with sensors that enable each individual to gather information about its local environment. However, agents cannot communicate with each other; this constraint enables testing of policies developed using the \gls{ctde} paradigm.

Under these conditions, the Rover Domain can be described as a \gls{modecpomdp} \cite{chatterjee2006markov,thakar2025multiagent}. The global state consists of the joint coordinate positions of the rover team. For every time step, each rover takes an action based on its local observations of the distance of \glspl{poi} and nearby rovers, and its current policy. An agent's action corresponds to a movement in the x and y directions up to a specified step maximum and to environmental dynamics. Activation of a \gls{poi} depends on (a) the number of rovers visiting it simultaneously, i.e., the coupling factor, and (b) the length of time they have been there. The global reward at each time step then depends on which \glspl{poi} have been successfully activated. \glspl{poi} may present different reward values depending on the experimental setup.

\subsection{Evolutionary Policy Optimization}
The \gls{nsgaii} algorithm was selected for its relevance to multi-objective optimization and its established use in evolutionary policy search \cite{ma2023comprehensive}. This makes it a natural baseline for evaluating the effects of entropy-influenced policy evaluation relative to conventional objective-space methods.

The policies evolved using \gls{nsgaii} are the weights and biases of neural network controllers for each agent. The parameters used are listed in Table \ref{tab:nsgaii_hyperparams}. Per the standard baseline \gls{nsgaii} algorithm, policies are evaluated primarily on their fitness as determined by the rewards gained for each objective in the Rover Domain. \gls{nsgaii} primarily preserves policies with the highest ranking, i.e., policies with fitnesses that are not dominated by any other policy's fitness on a single objective, and secondarily by crowding distance in the objective space.

However, Pareto ranking cannot decipher if policies have converged, or distinguish between policies that produce largely stagnate behavior and ones where agents are nearer to achieving higher rewards. Consequently, in a sparse reward environment, policies that achieve similar reward vectors but drastically different behaviors will be selected indiscriminately. Thus, the following section explains the novel introduction of entropy-influenced policy evaluation for multi-objective, multiagent joint policies.

\begin{table}[t]
\caption{Hyperparameters used in NSGA-II algorithm}
\label{tab:nsgaii_hyperparams}
\centering
\begin{tabular}{l c}
\toprule
Hyperparameter & Value \\
\midrule
Population Size & 75 \\
Generations & 10{,}000 \\
Policy Hidden Layers & [16, 16] \\
Mutation Rate & 0.75 \\
Mutation Scale & 0.5 \\
Weight Initialization Limit & 0.2 \\
Bias Initialization Limit & 0.2 \\
Entropy Neighbors ($k$) & 5 \\
Entropy Scale Factor ($\beta$) & $\{0.0, 0.05, 0.1, 0.25, 0.5\}$ \\
\bottomrule
\end{tabular}
\end{table}

\subsection{KNN Entropy-Augmented Policy Evaluation}
To prevent direct competition of task objectives with entropy, this work augments objective fitness scores with a scaled bonus of the measured entropy rather than introduce entropy as a competing objective. This method is intended to preserve the existing Pareto structure while promoting the discovery of more tradeoff solutions along the Pareto front.

A proxy for entropy, $\hat{H}$, is approximated using a \gls{knn}-based estimator \cite{pappas2021multiobjective} computed over joint trajectories within an episode:
\[\hat{H} = \frac{1}{N} \sum_{n=1}^{N} \log \epsilon_n
\] where $\epsilon_n$ denotes the distance from sample $n$ to its k\textsuperscript{th} nearest neighbor in the joint state space, and $N$ denotes the length of the episode. The augmented fitness score for any policy, $\pi$, of an objective, $i$, then becomes the approximated \gls{knn} entropy, $\hat{H}$ for the joint trajectories obtained using that policy, scaled by a factor, $\beta$, and added to the original fitness score, $f$:
\[
\tilde{f}_i(\pi) = f_i(\pi) + \beta \cdot \hat{H}(\pi), \quad \forall i \in {1, \ldots, m}
\]
By increasing $\beta$, policies that tend to exhibit broader exploration of the state space become more likely to be preserved during evolution. However, when $
\beta$ becomes too large, it can distort the true objective function. Ideally, $\beta$ values should be large enough to differentiate between similarly performing policies, without overshadowing true performance across objective dimensions the algorithm is optimizing for.

\subsection{Performance Metrics}
We seek to determine whether the addition of entropy as a secondary policy evaluation criterion can increase the overall hypervolume beneath the Pareto front. Hypervolume is computed after each evolutionary generation to characterize convergence and compare the  Pareto fronts obtained under different values of $\beta$.

\begin{figure}[h!]
  \centering

  \begin{subfigure}[b]{0.48\linewidth}
    \centering
    \includegraphics[width=\linewidth]{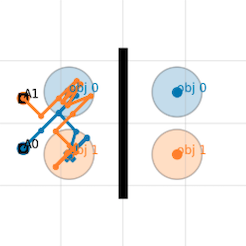}
    \caption{Agents favoring locally safe objectives}
    \label{fig:traj_a}
  \end{subfigure}
  \hfill
  \begin{subfigure}[b]{0.48\linewidth}
    \centering
    \includegraphics[width=\linewidth]{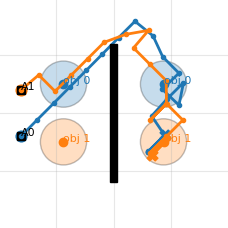}
    \caption{Agents discovering higher-value tradeoff solutions through exploration}
    \label{fig:traj_b}
  \end{subfigure}

  \caption{Example agent trajectories: (a) agents easily discover low value, repeatable \glspl{poi} and tend to adhere to this safe region; (b) agents may eventually discover higher value \glspl{poi} and therefore higher rewards on one or both objectives by accepting low rewards early on in the episode }
  \label{fig:wall}
\end{figure}

\begin{figure}[h!]
  \centering

  \begin{subfigure}[b]{0.48\linewidth}
    \centering
    \includegraphics[width=\linewidth]{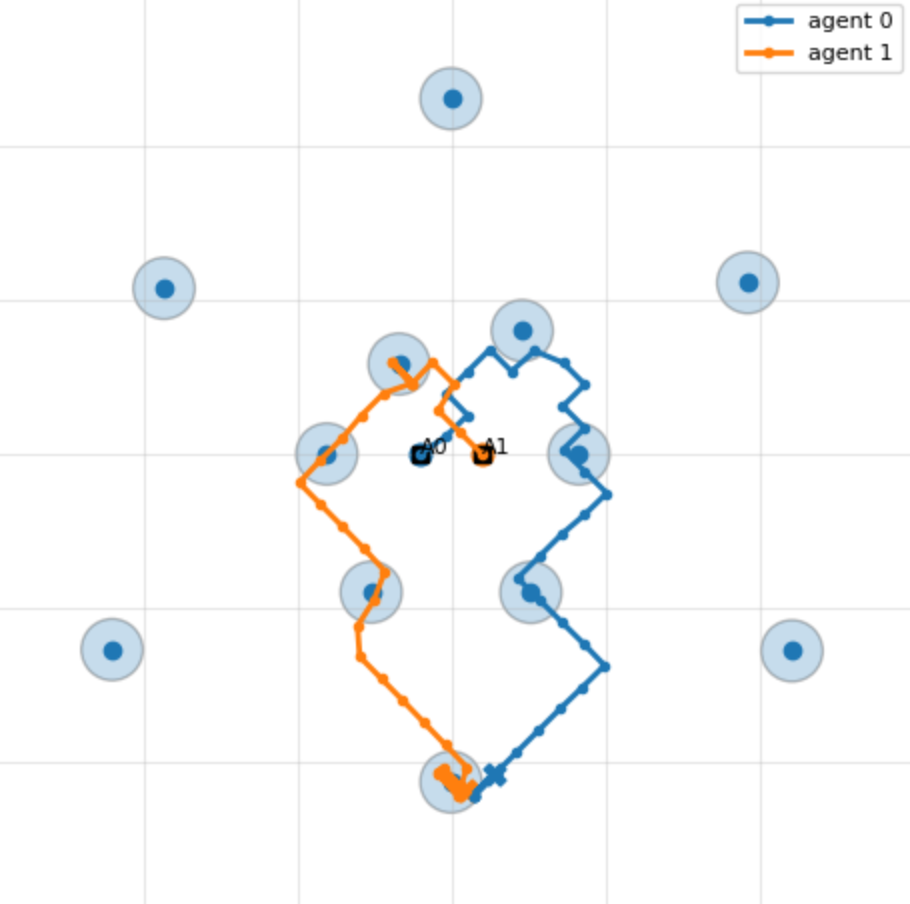}
    \caption{Agents optimizing for easy, fast rewards (inner ring)}
    \label{fig:time_a}
  \end{subfigure}
  \hfill
  \begin{subfigure}[b]{0.48\linewidth}
    \centering
    \includegraphics[width=\linewidth]{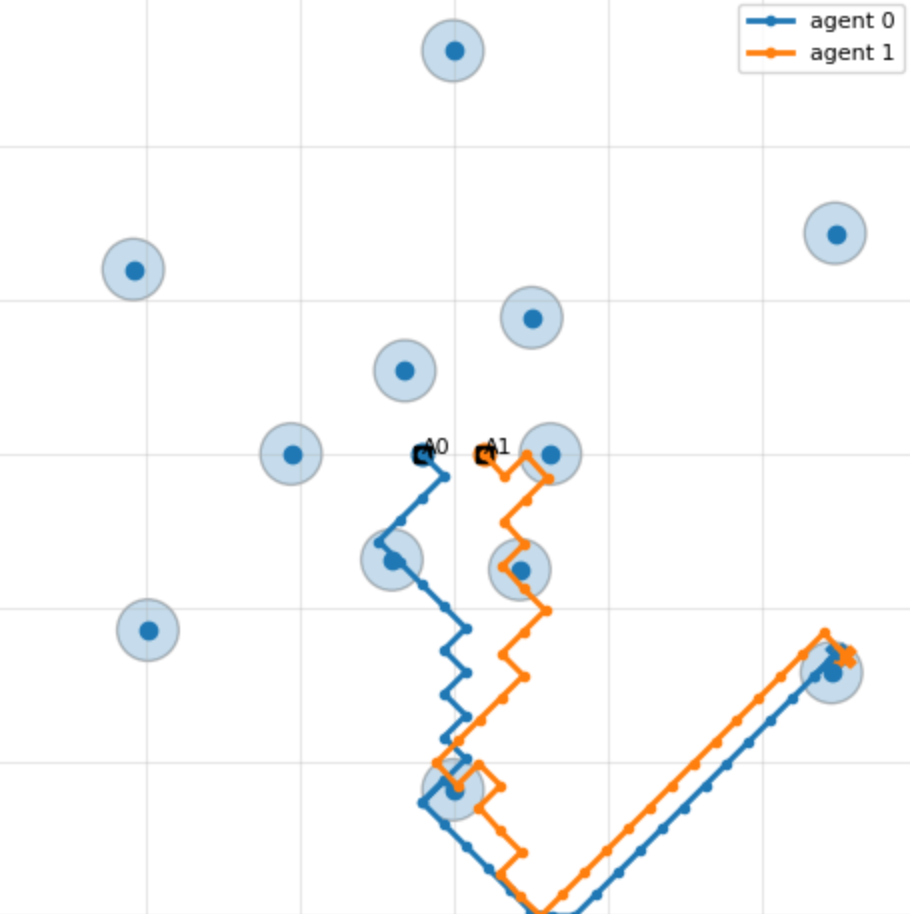}
    \caption{Agents trading delayed rewards for higher value \glspl{poi} (outer ring)}
    \label{fig:time_b}
  \end{subfigure}

  \caption{Example agent trajectories from the time experiment: (a) agents collect one-time rewards from close, 1-point \glspl{poi} before moving onto higher reward, 3-point \glspl{poi}; (b) agents collect close \glspl{poi} on the way to farther, higher reward \glspl{poi}}
  \label{fig:hypervolume}
\end{figure}

\section{Experiments}
In this section, we introduce the experiment setup in which the standard \gls{nsgaii} implementation is compared to the introduction of entropy-based bonuses to the policy evaluation procedure.

\subsection{Experiment Setups}
The purpose of this work is to investigate whether incorporating behavioral diversity into policy evaluation influences the diversity and quality of tradeoff solutions for teams of agents in a multi-objective setting.

\subsubsection{Experiment 1: Obstacle Navigation}
In this experiment, a two-rover team was tasked with maximizing rewards over two objectives; each \gls{poi} corresponded with one objective. The rovers were initially placed near low-value \glspl{poi}, in which they received one point for each time step they visited the POI concurrently. This sub-configuration of two neighboring \glspl{poi}, in which the agents must ``choose" between two mutually exclusive objectives, was configured in this manner to facilitate localized tradeoff exploration \cite{thakar2025multiagent}. However, an identical setup was then placed behind a simple wall, in which rovers must trade lower rewards in the beginning of the episode in order to achieve higher rewards near the end (5 points per \gls{poi} per time step). This discovery requires agents to explore beyond their immediate location despite the immediate rewards provided to them. This configuration is shown in Figure \ref{fig:wall}.

\subsubsection{Experiment 2: Temporal Tradeoff}
The purpose of this experiment was to evaluate temporal objectives. As opposed to \glspl{poi} representing different objectives, the \glspl{poi} in this experiment did not directly represent the objectives. Instead, the objectives were the reward collected across all \glspl{poi} over the first half of the episode, and then over the total episode. The agent team was rewarded one point each per near (inner ring) \gls{poi}, and three points each per outer \gls{poi}. Inner \glspl{poi} had a coupling factor of one, so agents could activate them individually. Outer \glspl{poi} were more difficult to activate, with a coupling factor of two. The episode length was short enough such that the most efficient team would be unable to collect all \glspl{poi} and be forced to make tradeoffs between collecting \glspl{poi} quickly in the episode, versus sacrificing easy, early rewards for higher value, resource intensive rewards.

\subsection{Compared Methods}
We compare the standard NSGA-II policy evaluation procedure with entropy-augmented policy evaluation based on scaled kNN entropy. Entropy bonuses represent an introduction of behavior-space influence in policy evaluation and are scaled using the parameter, $\beta$ = {0.05, 0.1, 0.25, and 0.5}. A $\beta$ value of 0.0 represents objective-space evaluation only.

\subsection{Analysis Protocol}
In the two-agent experiment, each trial was run for 10,000 time steps. Each $\beta$ value was run 30 times for Experiment 1, and 3 times for Experiment 2, using the same random seed for each $\beta$ value and a different random seed for each run. Experiment 2 should be interpreted as a preliminary evaluation due to the limited number of independent trials. All policies were rolled out at every timestep. The average and standard error of each beta value's hypervolumes over all runs and generations were collected and used to compare objective-space policy evaluation to a combined behavior- and objective-space evaluation. Hypervolume was selected as the primary performance metric because it jointly captures convergence and coverage of the Pareto front. For hypervolume calculation, a negative reference point was selected, i.e., (-1, -1) following standard hypervolume evaluation practice for this reward range.

\begin{figure}[h]
  \centering
  \includegraphics[width=\linewidth]{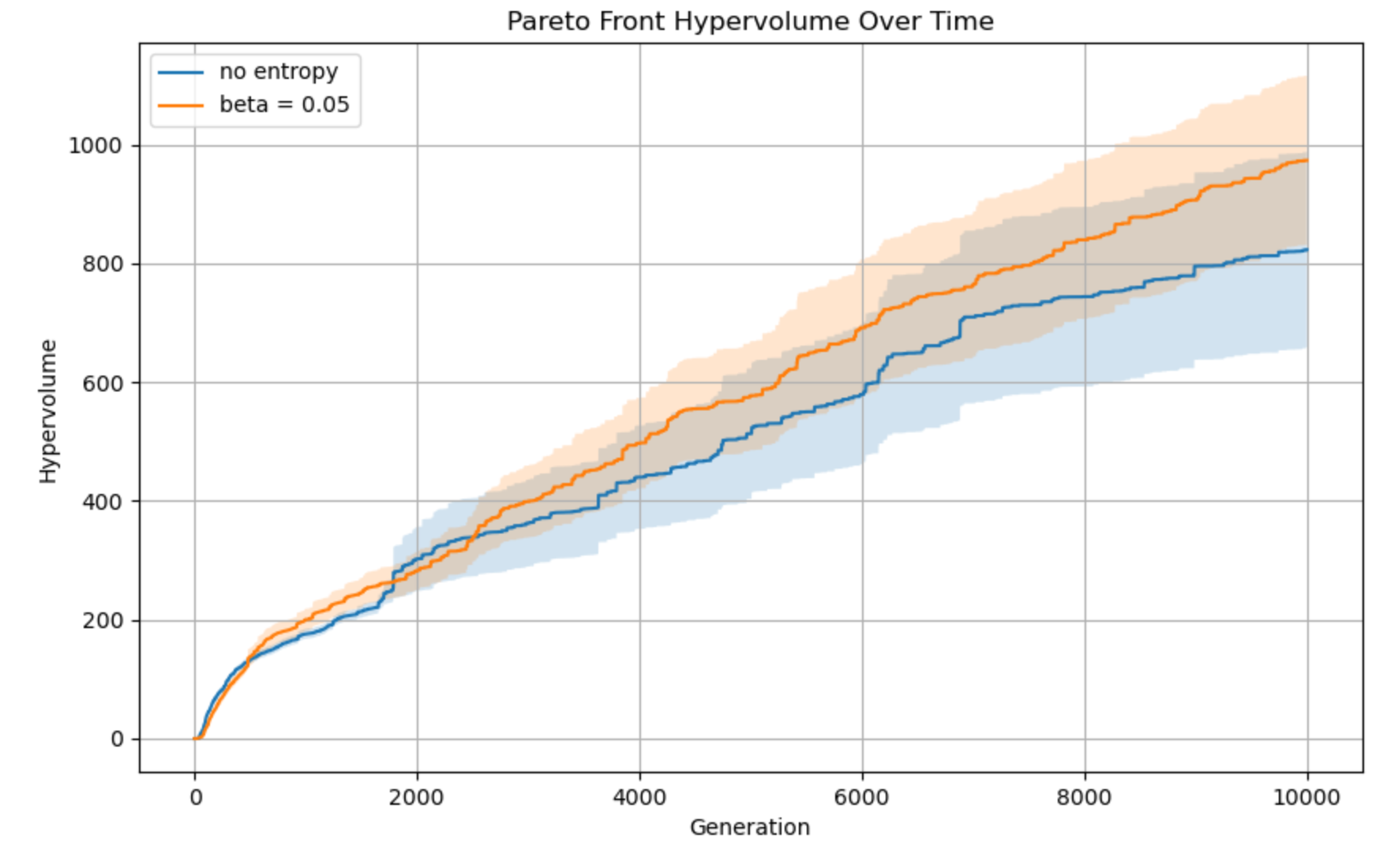}
  \caption{Experiment 1 Pareto front hypervolumes over \gls{nsgaii} generations with and without entropy-augmented fitness evaluation. Shaded regions represent the error.}
  \label{fig:3}
\end{figure}

\begin{figure}[h]
  \centering
  \includegraphics[width=\linewidth]{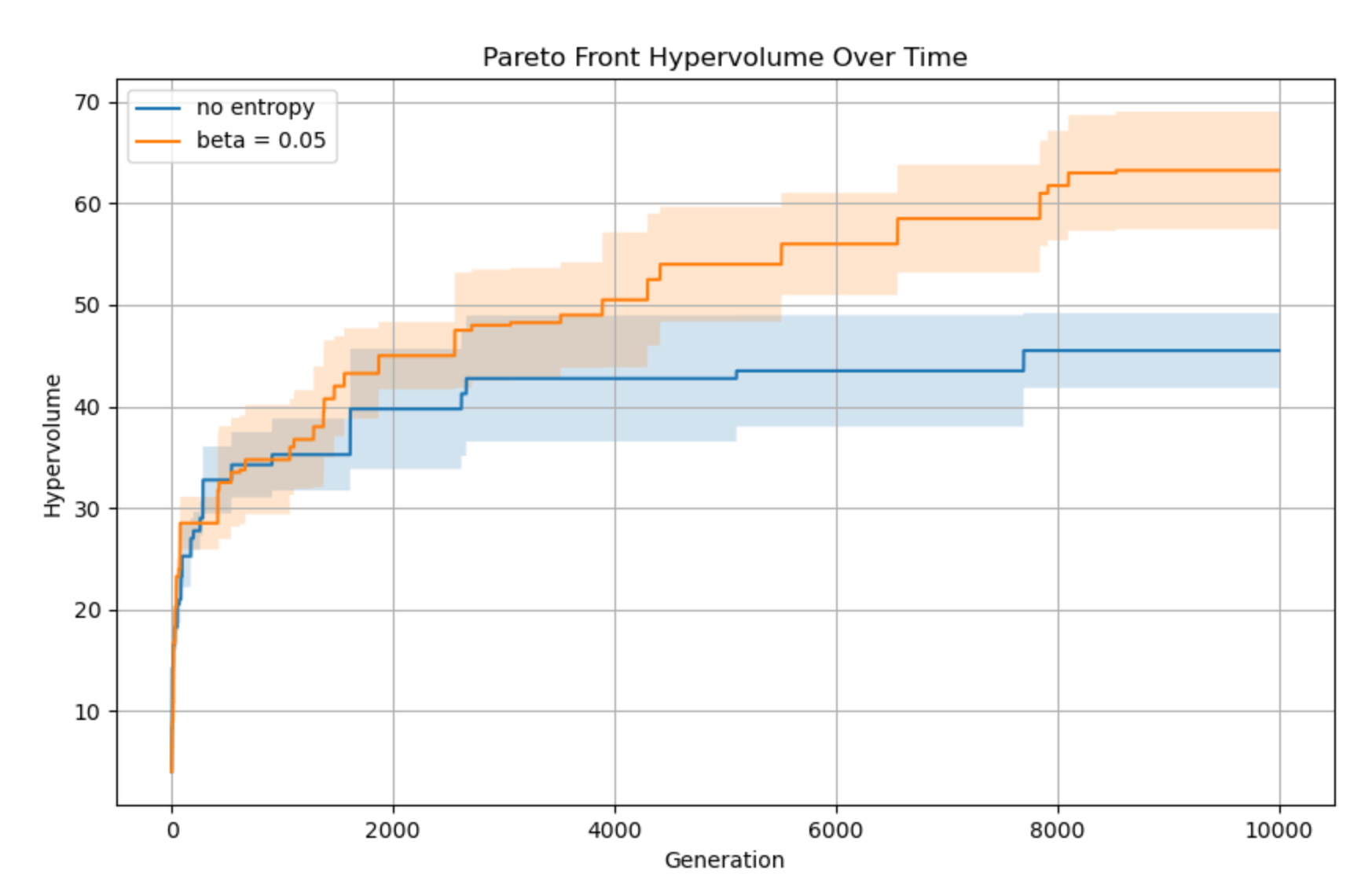}
  \caption{Experiment 2 Pareto front hypervolumes over \gls{nsgaii} generations, with shaded regions representing the error. The lower variance relative to Experiment 1 reflects the absence of a strong local optimum in the temporal tradeoff domain.}
  \label{fig:4}
\end{figure}

\section{Results and Discussion}
This section discusses the results from the study and evaluates the implications of introducing behavior-space awareness into the policy evaluation step of multi-objective evolutionary algorithms. The aim of this study is to evaluate whether entropy-augmented policy evaluation influences the quality of the Pareto fronts generated by NSGA-II. 

\subsection{Pareto Front Quality}
For evaluation, we observe the total hypervolume beneath each Pareto front, which indirectly measures both the quality and diversity of the optimal solution sets. Figures \ref{fig:3} and \ref{fig:4} compare the baseline with the best-performing entropy coefficient identified during parameter tuning. A broader range of $\beta$ values were tested in initial experiments (Table \ref{tab:nsgaii_hyperparams}); however, preliminary parameter tuning suggested that the small value of 0.05 provided the most promising results.

In Experiment 1, the entropy-augmented method achieved a hypervolume of $973.34 \pm 142$ compared to $822 \pm 164.79$ for the baseline (Table \ref{tab:exp1_hv}), an average increase of 18\%. However, as shown in Figure \ref{fig:3}, there is high variance in the results, causing uncertainty in the significance of the result. One possible explanation for the observed variance is that the wall creates a strong local optimum attractor as shown in Figure \ref{fig:wall} that heavily depends on early random initialization and if the agents happen to explore around the edge simultaneously. Therefore, to reach statistically significant results, this experiment likely requires many more trials to reduce the standard error. However, despite the high variance, entropy augmentation produced higher hypervolume than the baseline in most runs, suggesting that entropy augmentation may be beneficial even in high-variance settings.

In Experiment 2, the entropy-augmented method achieved a final hypervolume of $63.25 \pm 5.79$, compared with $42.67 \pm 3.33$ (Table \ref{tab:exp1_hv}). Although these preliminary experiments showed a substantial improvement in average hypervolume, the evaluation was conducted using a limited number of trials. Consequently, these results should be interpreted as exploratory and warrant additional evaluation to determine their robustness.

\begin{table}[t]
\centering
\caption{Final Pareto-front hypervolume for Experiment 1 across entropy coefficients.
Values are reported as mean $\pm$ standard error (SE); n=29.}
\label{tab:exp2_hv}
\begin{tabular}{c c}
\toprule
$\beta$ & Hypervolume \\
\midrule
0.0  & $822 \pm 164.79$ \\
\textbf{0.05} & $\textbf{973.34} \pm \textbf{142.43}$ \\
\bottomrule
\end{tabular}
\end{table}

\begin{table}[t]
\centering
\caption{Final Pareto-front hypervolume for Experiment 2 across entropy coefficients.
Values are reported as mean $\pm$ standard error (SE); n=3.}
\label{tab:exp1_hv}
\begin{tabular}{c c}
\toprule
$\beta$ & Hypervolume \\
\midrule
0.0  & $42.67 \pm 3.33$ \\
\textbf{0.05} & $\textbf{63.25} \pm \textbf{5.79}$ \\
\bottomrule
\end{tabular}
\end{table}

\section{Conclusions}
To our knowledge, this is the first work to incorporate both objective- and behavior-space diversity into Pareto optimization for multiagent systems. Rather than introducing entropy as a competing objective, an entropy-based bonus is distributed directly across objective fitness values, preserving the global Pareto structure. Preliminary results suggest that incorporating behavioral diversity into policy evaluation is a promising direction for improving multi-objective evolutionary search. However, additional evaluation across a broader range of random seeds, domains, and problem settings is needed to assess the robustness and generality of the observed improvements.

\subsection{Limitations and Future Work}
While the proposed fitness shaping strategy is simple to incorporate into existing \glspl{moea}, several challenges remain. The efficacy of the method depends on the relative magnitude of the entropy bonus. If the entropy bonus becomes too large relative to the task rewards, it can dominate the fitness evaluation and diminish the influence of the original objectives. Selecting or adapting the scaling factor remains an important challenge, particularly in sparse-reward domains.

Future work will explore more complex environment configurations, such as incorporating additional agents and \glspl{poi}, as well as extending episode length to better evaluate the discovery of novel behaviors. In addition, we will investigate the conditions under which entropy augmentation is beneficial, including its sensitivity to problem characteristics, random initialization, and the choice of behavioral characterization. While entropy was selected as an initial implicit summary of team behavior, alternative behavioral summaries and diversity measures may better capture task-relevant coordination patterns while preserving diverse collaborative behaviors during evolutionary search.


\bibliographystyle{named}
\bibliography{ijcai26}

\end{document}